\documentclass[a4paper,fleqn,usenatbib]{mnras}

\usepackage[T1]{fontenc}
\usepackage{ae,aecompl}
\newcommand\be{\begin{equation}}
	\newcommand\ee{\end{equation}}
\newcommand\bea{\begin{eqnarray}}
	\newcommand\eea{\end{eqnarray}}

\newlength{\wdth}

\usepackage{graphicx}	

\usepackage{amsmath}	
\usepackage{amssymb}	
\usepackage{enumerate,multirow}

\title[AGN heating of the ICM] 
{Heating of the intracluster medium by buoyant bubbles and sound waves}
\author[Iqbal et~al.]
{Asif Iqbal$^{1}$\thanks{
asif.ahangar@cea.fr}, Subhabrata Majumdar$^{2}$\thanks{subha@tifr.res.in} Biman B. Nath$^{3}$\thanks{
biman@rri.res.in}, , Suparna Roychowdhury$^{4}$\thanks{suparna@sxccal.edu} \\
  $^{1}$Universit\'e Paris-Saclay, Universit\'e Paris Cit\'e CEA, CNRS, AIM, 91191, Gif-sur-Yvette, France\\
  $^{2}$Tata Institute of Fundamental Research, 1 Homi Bhabha Road, Mumbai, 400005, India\\ 
  $^{3}$Raman Research Institute, Sadashiva Nagar, Bangalore, 560080, India\\
  $^{4}$Department of Physics, St. Xavier's College, 700016, Kolkata, India
}
\date{Submitted to MNRAS}
\pubyear{2022}

\begin{document}
\label{firstpage}
\pagerange{\pageref{firstpage}--\pageref{lastpage}}
\maketitle
\begin{abstract}
Active galactic nuclei (AGN)  powered by the central Super-Massive Black Holes (SMBHs) play a major role in modifying the thermal properties of the intracluster medium (ICM). In this work, we implement two AGN heating models: (i) by buoyant cavities rising through stratified ICM (effervescent model) and, (ii) by viscous and conductive dissipation of sound waves (acoustic model). Our aim is to determine whether these heating models are consistent with ICM observables and if one is preferred over the other. We assume an initial entropy profile of ICM that is expected from the purely gravitational infall of the gas in the potential of the dark matter halo. We then incorporate heating, radiative cooling, and thermal conduction to study the evolution of ICM over the age of the clusters. Our results are:  (i) Both the heating processes can produce comparable thermal profiles of the ICM  with some tuning of relevant parameters. (ii) Thermal conduction is crucially important, even at the level of 10\% of the Spitzer values,  in transferring the injected energy beyond the central regions, and without which the temperature/entropy profiles are unrealistically high. (iii) The required injected AGN power scales with cluster mass as $M_{\rm vir}^{1.5}$ for both models. (iv) The required AGN luminosity is comparable with the observed radio jet power, reinforcing the idea that AGNs are the dominant heating source in clusters. (v) Finally, we estimate that the fraction of the total AGN luminosity available as the AGN mechanical luminosity at $0.02r_{500}$ is less than 0.05\%.
\end{abstract}
\section{Introduction}
Current and future X-ray and CMB missions (like eROSITA, Athena, Simons Array, CMB-S4, CMB-HD, etc) have cluster cosmology and cluster physics as two of their main drivers. The synergy of cosmology and cluster gas physics, intertwined through the nature of the ICM, lies at the core of realizing the science goals.
The physics of the ICM is complex due to the multiple energetic physical processes, having both temporal and spatial dependence, involved in it. With the advent of the current X-ray satellites, Chandra, XMM-Newton and eROSITA, it is now believed  that the energetics of the ICM is regulated by heating from non-gravitational sources like AGN and SNe in galaxies, in addition to the heating at the accretion shock due to gravitational collapse \citep{White-Rees_1978} and radiative cooling.  One of the most important implications of these observations is that the central gas must experience some kind of heating plausibly due to the same feedback mechanism that prevents cool cores from establishing  significant cooling flows that were predicted by earlier, low-resolution, X-ray observations  (see \cite{Fabian_94,Peterson_01,Peterson_06} and references therein). Establishing the source of this heating, and understanding when and how it takes place, has become a major topic of study in extra-galactic astrophysics. In addition to the cooling flow problem, another important issue that came into focus recently  is the existence of an enhancement in the entropy profile  within  the core ($<$ 100 kpc) of the cluster (see \citet{10} and references therein). This entropy enhancement is found to be more pronounced in non cool-core (NCC) clusters  compared to the cool-core  (CC) clusters.

The complexities of ICM also manifest in the so-called ``cluster scaling relations''. The theory of hierarchical structure formation  predicts cluster scaling relations to be self-similar \citep{Kaiser1986, Sereno2015}. However, observations show departure from self-similarity; for example, the luminosity-temperature ($L_x-T$) relation for self-similar models predict a shallower slope ($L_x\propto T^{2}$) than observed ($L_x\propto T^{3}$)  \citep{Pratt2009}. Similarly, Sunyaev-Zel'dovich (SZ) scaling relations also show similar departure \citep{Holderb2001,Andrade-Santos2021}.    

Several processes have been proposed to explain the observations: pre-heating of the infalling gas due to early feedback processes in high-redshift galaxies \citep{babuletal02}, AGN feedback from quasars or radio jets \citep{binney&tabor95, rephaeli&silk95, Nath2002}, conduction of thermal energy from the outer shock-heated regions carried by electrons \citep{voigt&fabian04,rasera&chandran08}, and gas sloshing from minor and major mergers \citep{fabian&daines91}. While the verdict is still out for early pre-heating and thermal 	conduction, the ability of AGN feedback to stem cooling flows, and to break self-similarity in scaling relations,  has  been demonstrated in several hydrodynamical simulations \citep{sijacki&springel06, khalatyanetal08, puchweinetal08, fabjanetal10, duboisetal10, mccarthyetal10, teyssieretal11}.  It seems, therefore, natural to consider such an AGN feedback mechanism as a key ingredient to account for the excess energy or entropy in the ICM.  However, one still needs to understand the exact physical process that helps evolve the excess entropy with time and distance from the SMBHs powering the central AGN.

In earlier work,  \cite{Iqbal2017a,Iqbal2017b} found that the presence of non-gravitational energy per particle, related to excess entropy beyond $r_{500}$ is almost negligible,  thereby ruling out pre-heating models at a large confidence level. Subsequently, \cite{Iqbal2018} showed that AGN feedback and radiative cooling are jointly responsible for the state of the ICM in the central regions, $\textrm{r} \lesssim 0.3\textrm{r}_{500}$. Similarly, \cite{Gaspari2014} showed that AGN feedback can naturally regulate the thermodynamical state of ICM up to $\textrm{r}\approx0.2\textrm{r}_{500}$.   Given the importance of AGN feedback and radiative cooling in the inner regions and the lack of excess energy in the outer regions, it is natural to investigate the radial dependence of the feedback energetics.

There are a number of models and simulations have been developed to explain the AGN feedback, however, their validity and applicability are mitigated by the implicit assumptions used in these models. The issue of the physical mechanism of heating remains elusive precisely for these reasons, despite the plethora of models (see, for example,  \cite{Soker2022}  who list 7 main  AGN heating models).  
In this work, we consider two models of mechanical heating by central AGN and ignore the other heating processes. One concerns the  work done by  bubbles (cavities) blown by the AGN jet and carried towards the outer regions of the cluster by the pressure gradient in the ICM \citep{Ruszkowski2002,Roychowdhury2004}. The other model explores the possibility of heating via viscous dissipation and thermal conduction of the energy by sound waves generated by some other phenomena related to the jet \citep{Fabian2005, Zweibel2018}.  These two models have been analytically worked out in detail, with heating rates written down in analytical forms, and hence can be compared to X-ray and SZ observations.  It is not a priori clear whether or not the acoustic and the effervescent heating both satisfy the observations. And if they do so, it is important to determine for what values of the parameters they are valid.  

In this paper, we compare these two modes of energy deposition into the ICM by AGN, combined with radiative cooling and thermal conduction, for clusters of different masses.  We trace  the  evolution of the thermal properties of the ICM over its lifetime. Given that the ICM is not affected by feedback far from the core, we parameterize feedback models such that they affect the thermal structure of ICM up to $0.1\textrm{r}_{500}$ and  $0.3\textrm{r}_{500}$. Finally, we estimate the relation between the mechanical energy injected by the AGN and the cluster mass and compare it with scaling relations derived from the complementary observations. For simplicity, we ignore convection  and  cooling flows in this work. Both these effects are expected to be significant below 0.1$r_{500}$ and we plan to consider these effects in a companion paper. 

Throughout this work, we adopt a cosmology with $H_{\rm 0}=70$ km s$^{-1}$ Mpc$^{-1}$, $\Omega_{\rm m}=0.3$ and $\Omega_{\rm \Lambda} =0.7$.  Further, 
$E(z)$ is the  is the ratio of the Hubble constant at redshift $z$ to its present value, $H_{\rm 0}$ and $h_{\rm 70}=H_{\rm 0}/70=1$.

\section{Cluster Model}
\subsection{The dark matter profile}
We work with the Navarro-Frenk-White (NFW) density profile  ($\rho_{\textrm{tot}}$) \citep{Navarro1996, Navarro1997} of galaxy clusters  given by
\begin{equation}
	\rho_{\textrm{tot}}(r)=\frac{\rho_s}{y(1+y)^2},
	\label{darkmater}
\end{equation}
where $y=r/r_{\rm s}$, $r_{\rm s}$ is the scale radius and $\rho_{\rm s}$ is the normalization of the density profile.  The total mass profile  ($M_{\rm tot}$) of galaxy clusters can then be simply expressed as 
\begin{equation}
	M_{\rm tot}=4\pi r^3_{\rm s}\rho_{\rm s}\left[\ln(1+y)-\frac{1}{1+y}\right]
	\label{eqA}
\end{equation}	
For a given total  virial mass of cluster ($M_{\rm vir}$),  the virial  radius, $R_{\rm vir}(M_{\rm vir},z)$ is found using
$R_{\rm vir}=\left[\frac{M_{\rm vir}}{4\pi/3\Delta_{\rm c}(z)\rho_{\rm c}(z)}\right]^{1/3}$ 
\citep{Peebles1980}, where the overdensity 
$\Delta_{\rm c}(z)=18\pi^2+82(\Omega_{\rm m}(z)-1)-39(\Omega_{\rm m}(z)-1)^2$ \citep{Bryan1998}.
 The concentration parameter is related to the  $r_{\rm s}$ as  $c_{\rm vir}=R_{\rm vir}/r_s$  where $R_{\rm vir}$ is the virial mass.  Numerical simulations predict self-similar relation between $c_{\rm vir}-M_{\rm vir}$ and we adopt the expression for the concentration parameter from \cite{Duffy2008} 
\begin{equation}
	c_{\rm vir} = 7.85\,\left(\frac{M_{\rm vir}}{2\times10^{12}\,h^{-1} M_{\odot}}\right)^{-0.081}\, \left(1+z\right)^{-0.71}
\end{equation}
The $c_{\rm vir}-M_{\rm vir}$ relation from \cite{Duffy2008} has been found  to be consistent with the Subaru  weak lensing estimates of \cite{Okabe2010}.

\subsection{The fiducial ICM profile}
Numerical simulations, backed by current X-ray and SZ observations, show that the ICM pressure profile follows a universal  form  which is well described by a generalized NFW
model \citep{Nagai2007,11,Planck2013V}
\begin{equation}
	\dfrac{P_g(x)}{P_{\rm 500}}  =  \dfrac{P_{0}}{(c_{500}x)^{\gamma} [1+(c_{500}x)^{\alpha}]^{(\beta-\gamma)/\alpha}},
	\label{gnfw}
\end{equation}
where $x=r/r_{500}$. $P_0$, $c_{500}$, $\gamma$, $\alpha$, $\beta$ are the model parameters and
\begin{eqnarray}
	P_{\rm 500}=1.65\times 10^{-3} E(z)^{8/3}\qquad \qquad\qquad \qquad \nonumber\\
	\quad\quad  \quad  \times \left[\frac{M_{500}}{3\times10^{14} h_{70}^{-1}M_{\odot}\,}
	\right]^{2/3}\, h_{70}^{2} \textrm{keV cm$^{-3}$}.
	\label{e:nor}
\end{eqnarray}
$P_{\rm 500}$ reflects the self-similar dependence with mass and redshift. Moreover,  simulations have shown no significant evolution outside of the cluster core \citep{Battaglia2012,Planelles2017}, which has been also confirmed observationally  \citep{McDonald2014,Adam2015}.
In the present work, we consider   \cite{Planelles2017} best fit non-radiative  pressure profile  ($P_0=6.85$, $c_{500}=1.09$,  $\gamma=0.31$, $\alpha=1.07$ and $\beta=5.46$) as our baseline pressure profile. It is worth mentioning here that the  \cite{Planelles2017}  did not find additional mass dependence of pressure profile like in \cite{11}.  Given an initial non-radiative pressure  (i.e Eq.~(\ref{gnfw})) and NFW model  for total mass (i.e Eq.~(\ref{eqA})), the density ($\rho_{\rm g}$) profile (and hence temperature ($T_{\rm g}$)) of the ICM can be determined using hydrostatic equation
\begin{equation}
\rho_{\rm g}(r)=\frac{r^2}{G M_{\rm tot}( <r ) }\frac{dP_{\rm g}}{dr}.
	\label{H:he2}
\end{equation}
 Finally, note that the the overall conclusion of this work is independent of the choice of the initial ICM  profile used, for example using non-radiative profile from \cite{Voit2005} would have made no significant change.

\subsection{Central AGN heating}
Here we discuss briefly the two models of mechanical heating by central AGN: the acoustic model and the effervescent model. 

\subsubsection{Effervescent heating model}
The central AGN is responsible for 
inflating buoyant bubbles of relativistic plasma in the ICM in the effervescent heating model \citep{Churazov2001, Begelman2001, Ruszkowski2002, Roychowdhury2004}. The timescale for bubbles to cross the cluster, which is of the order of the free-fall time, is found to be shorter than the cooling timescale. It is assumed that the number flux of bubbles is large such that the flux of bubble energy through the ICM approaches a steady state. This, in turn, implies that the details of the energy injection process such as the number flux of bubbles, bubble radius, filling factor and the rate of rise do not affect the average	heating rate. 

We assume that that the relativistic gas inside the bubble does not mix with the ICM very efficiently and that bubbles push aside the X-ray emitting gas, thus excavating depressions in the ICM which should be detectable as apparent cavities in the X-ray images. Indeed, this scenario is vindicated through  Chandra and XMM-Newton observations which have seen cavities  far away from the central regions of the cluster \citep{Shin2016}.  In this scenario, the bubbles can expand and do $pdV$ work on the ambient medium, as they rise in the cluster pressure gradient, thus converting the internal energy of the bubbles to thermal energy of the ICM within a pressure  scale height of where it is generated.  It is important to mention, although bubbles have been detected out to large radii, some 3D hydrodynamical  simulations have shown a strong mixing of the bubbles with the ICM  \citep{Hillel2020}. In such cases, effervescent heating might actually be a subdominant process.

In steady state (assuming spherical symmetry) and assuming negligible mixing, the energy flux carried by the bubbles, during adiabatic bubble inflation, is given by  \citep{Begelman2001,Roychowdhury2004}
\be
F_{\rm \scriptscriptstyle b} \propto \frac{P_{\rm \scriptscriptstyle b}(r)^{(\gamma_{\rm
			\scriptscriptstyle b}-1)/\gamma_{\rm \scriptscriptstyle b}}}{r^{2}}
\ee
where $P_{\rm \scriptscriptstyle b}(r)$ is the partial pressure of relativistic buoyant	gas inside the bubbles at cluster radius $r$ and the relativistic adiabatic index of buoyant gas $\gamma_{\rm \scriptscriptstyle b} =4/3$. 	Assuming that the partial pressure inside these bubbles scales as the thermal pressure of the ICM, the volume heating rate $\epsilon_{\rm heat}(r)$ can be expressed as \citep{Begelman2001,Roychowdhury2004}
\bea
\epsilon_{\rm heat}(r) &\approx& r^{2}h(r) {\mathbf {\nabla}}\cdot ({\mathbf {\hat
		r}}F_{\rm \scriptscriptstyle b})\nonumber \\
&=& h(r)P_{\rm \scriptscriptstyle g} ^{
	(\gamma_{\rm \scriptscriptstyle b}-1)/\gamma_{\rm \scriptscriptstyle b}}
{1 \over r}{{d \ln P_{\rm \scriptscriptstyle g}} \over {d \ln r}}
\label{eq:heatingrate}
\eea
where  $h(r)$ is given by 
\begin{equation}
	h(r) = {L^{\rm inj}_{\rm Eff} \over {4 \pi r^2}}[1-\exp(-r/r_
	{\rm \scriptscriptstyle 0})]\exp(\rm -r/r_
	{\rm \scriptscriptstyle cutoff})\, q^{-1} .
	\label{eq:norm_finc}
\end{equation}
In Eq.~\ref{eq:norm_finc}, $L^{\rm inj}_{\rm Eff}$ is the time-averaged energy injection rate, $r_{\rm \scriptscriptstyle 0}$ represents the  transition  from bubble formation region  to the buoyant (effervescent) phase and $r_{\rm \scriptscriptstyle {cutoff}}$  is the outer heating cutoff radii. The term  $h(r)$ thus takes into account the fact that the volume heating rate is maximum near the inner cut-off radius and falls off exponentially with increasing radius. In our calculations, we fix $r_{0}$  to be equal to $0.015r_{500}$.  We note that our final results are not sensitive to the choice of  $r_{\rm \scriptscriptstyle 0}$.
The normalization factor $q$ is defined by 
\begin{equation}
	q=\int_{r_{\rm \scriptscriptstyle ini}}^{r_{\rm \scriptscriptstyle max}}
	P_{\rm g} ^{(\gamma_{\rm
			\scriptscriptstyle b}-1)
		/\gamma_{\rm \scriptscriptstyle b}}{1 \over r}{{d \ln P_{\rm g} } \over {d \ln r}}
	[1-\exp(-r/r_
	{\rm \scriptscriptstyle 0})]\exp(\rm -r/r_
	{\rm \scriptscriptstyle cutoff})\, dr
\end{equation}
where we fix $r_{\rm{max}}=R_{\rm vir}$.

\subsubsection{Acoustic heating model}
In the acoustic heating model \citep{Fabian2005,Fabian2016,Yang2016},  the ICM is heated through the dissipation of adiabatic acoustic waves produced from the central AGN. It has been shown through hydrodynamical simulations \citep{Sternberg2009} that the bubbles can also excite several consecutive sound waves without the need to invoke periodic jet launching episodes.  Assuming the average acoustic luminosity 
$L^{\rm inj}_{\rm Aco}$, injected into the ICM at $r_{0}$, the acoustic luminosity  surviving a given radius $r$ given by $L_{\rm Aco}$ will depend on the dissipation length  ${\ell_{\rm Aco}}$  (i.e $\frac{d L_{\rm Aco}}{dr}=-\frac{L_{\rm Aco}}{\ell_{\rm Aco}}$)  of the ICM as  \citep{Fabian2005}
\begin{equation}
	L_{\rm Aco}(r)=L^{\rm inj}_{\rm Aco}\times \exp\left(-\int_{r_{0}}^r\frac{1}{\ell_{\rm Aco}}\,dr\right).
\end{equation}
As before, we fix  $r_{0}$  to be $0.015r_{500}$.	 Assuming that heating is  due to kinematic viscosity ($\nu$) and thermal conductivity ($\kappa$), the acoustic dissipation length in the ICM can be written as \citep{Fabian2005}
\begin{equation}
	\ell_{\rm Aco}(r)=697\,\frac{n_{\rm e}(T_7)^{-1}(f_{-6})^{-2}}{\left(\frac{\xi_\nu}{0.1}\right)+11.8\left(\frac{\xi_\kappa}{0.1}\right)}{\rm kpc}
	\label{dissipationlength}
\end{equation}
where $f_{\rm -6}$ is the frequency of the sound wave in the units of mega year ($f_{-6}=f/(10^{-6}{\rm yr}^{-1})$, $T_{\rm7}$ is the temperature of the ICM in the units of $10^{7}$ K  ($T_{7}=T/10^{7}$) and $n_{\rm e}$ is the electron number density in cm$^{-3}$.  $\xi_\nu$ and $\xi_{\kappa}$ represent the viscosity and conduction fractions respectively of their Spitzer values in the absence of a magnetic field
\begin{eqnarray}
	\nu&=&1.0\times 10^{25}T_7^{5/2}n_{\rm e}^{-1}\xi_\nu  \nonumber \\
	\frac{\kappa}{\rho_{\rm g} c_{\rm p}}&=&2.36\times 10^{26}T_7^{5/2}n_{\rm e}^{-1}\xi_{\kappa}
\end{eqnarray}
where  $\rho_{\rm g} $ is the gas density and  $c_{\rm p}$ is the specific heat at constant pressure. The volume heating rate due to viscous and conductive dissipation is then given by
\begin{equation}
	\epsilon_{\rm heat}(r)=\frac{L_{\rm Aco}(r)}{4\pi r^2\ell_{\rm Aco}}.
\end{equation}
Applying this idea to the Perseus cluster, \cite{Fabian2005} suggested energy dissipation due to frequencies in the range $f_{-6}=0.2-1$, with the slope $\zeta=1.8$, to balance the radiative cooling at the cluster cores. In this work, we will consider heating by sound waves with $\zeta=1.8$ such that  acoustic luminosity in a frequency interval ($f$, $f+df$) is given by
\begin{equation}
	L^{\rm inj}_{\rm Aco,spec}(f)=A_{\rm norm}f^{-\zeta}
\end{equation}
where $A_{\rm norm}$ sets the normalization such that total acoustic injected luminosity $L^{\rm inj}_{\rm Aco}$ is given by $L^{\rm inj}_{\rm Aco}=A_{\rm norm}\,\int L^{\rm inj}_{\rm Aco,spec}(f) df$.  However, we will consider different frequency ranges, depending on the radial extent  of feedback,  suitable for our analysis.
The  modified volume heating rate can then be written as
\begin{equation}
	\epsilon_{\rm heat}=\int \frac{L^{\rm inj}_{\rm Aco,spec}(f)}{4\pi r^2\ell_{\rm Aco}}\exp\left( -\int_{r_{\rm ini}}^r\frac{1}{\ell_{\rm Aco}}\,dr\right)\,df.
\end{equation}
Note, that the higher frequencies will produce a higher heating rate but are confined to a smaller region as opposed to lower frequencies which will produce relatively less heating but up to a larger area. The total heating is frequency averaged over the spectrum in the spectral range taken for a particular cluster. 

\begin{figure*}
	\centering	
	\includegraphics[width=0.96\textwidth]{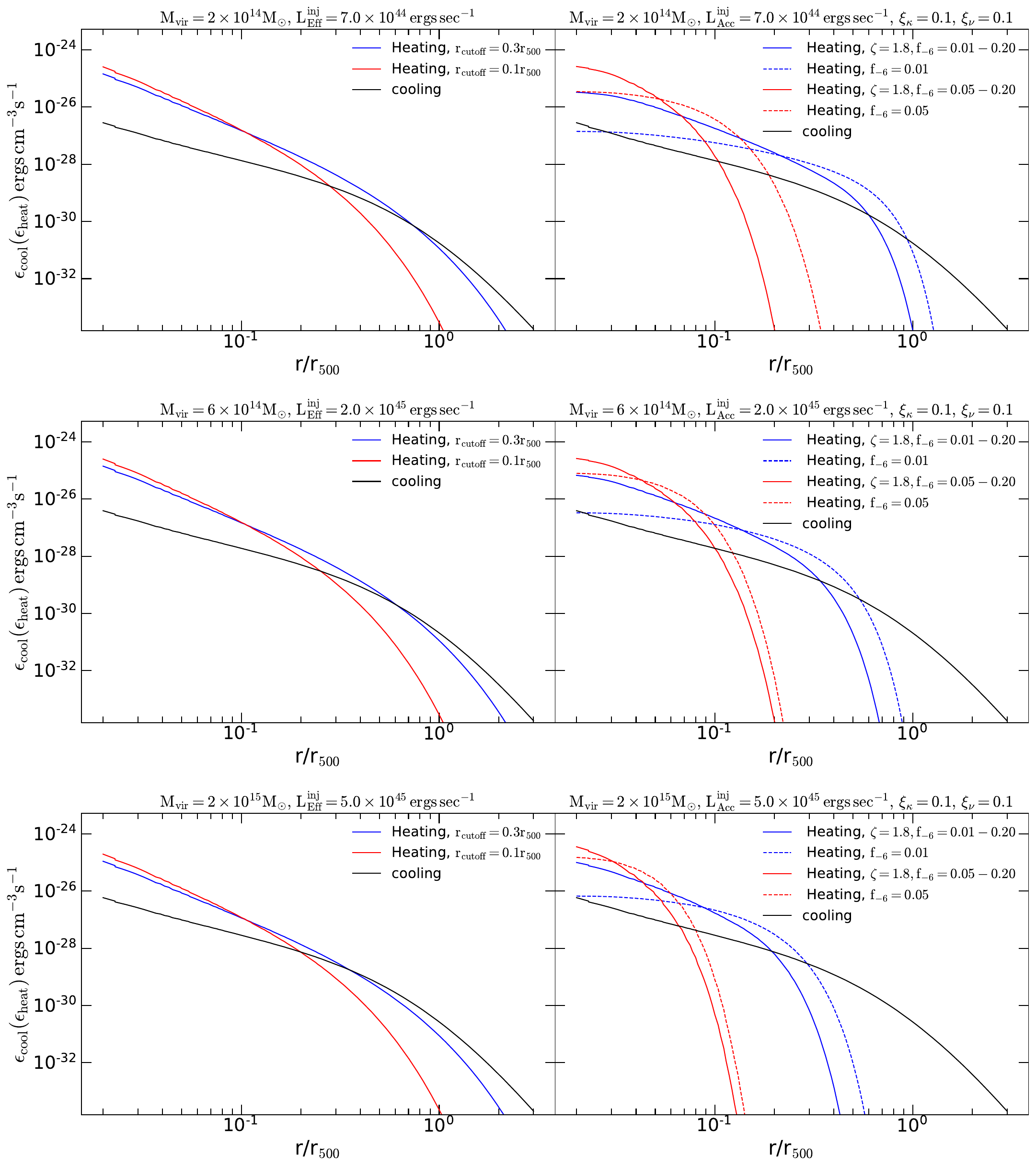}   
	\caption{
		Comparing cooling and heating rate for effervescent and acoustic models for three different cluster masses. Left panels: Initial cooling rate  (black line) versus heating rate  (blue and red lines) in effervescent case for  $2\times10^{14}M_{\rm \odot}$ (top),  $6\times10^{14}M_{\rm \odot}$ (middle) and  $2\times10^{15}M_{\rm \odot}$ (bottom) clusters at $z=0$.  Right panels:  Initial cooling rate (black line) versus heating rate with $\xi_{\kappa}=0.1$ and  $\xi_{\nu}=0.1$ (blue and red  lines) in acoustic case for  $2\times10^{14}M_{\rm \odot}$ (top),  $6\times10^{14}M_{\rm \odot}$ (middle) and  $2\times10^{15}M_{\rm \odot}$ (bottom) clusters at $z=0$.  Note that in case of acoustic heating, solid lines are obtained by assuming  a  frequency spectrum of $\zeta=1.8$ and  frequency range of $f_{-6}=0.01-0.20$ (solid blue line) $f_{-6}=0.05-0.20$ (solid red line) while as  dashed lines represent  heating profiles for  single frequency of $f_6=0.01$ (dashed blue line) and   $f_{-6}=0.05$ (dashed red line).
	} 
	\label{fig1}
\end{figure*}

\subsection{Radiative cooling and Conduction}
In the case of galaxy clusters, radiative cooling is dominated by free-free emission. The emissivity per unit volume can be expressed as
\begin{equation}
	\epsilon_{\rm cool} = n_e^2 \Lambda_N \frac{\mu_e}{\mu_h}\, {\rm erg\, s^{-1} \, cm^{-3}}
\end{equation}
where $\mu_h=1.26$. We consider the cooling function `$\Lambda_N$' from \cite{Tozzi2001} given by
\begin{equation}
	\Lambda_N = C_1 (kT)^\alpha + C_2 (kT)^\beta +C_3
	\label{lambdan}
\end{equation}
where  $\alpha=-1.7$ and $\beta= 0.5$. The constants $C_1=8.6\times 10^{-25}$ erg cm$^3$ s$^{-1}$ keV$^{-\alpha}$, $C_2=5.8\times 10^{-24}$  erg cm$^3$ s$^{-1}$
keV$^{-\beta}$ and $C_3=6.3\times 10^{-24}$  erg cm$^3$ s$^{-1}$ are for metallicity of $0.3$ $Z_\odot$.

In the presence of a thermal gradient, the heat flux due to thermal conduction is given by
\begin{equation}
	{\rm F}_{\rm cond} = - \kappa \nabla T.
\end{equation}  
One can easily see from the above equation that the Spitzer thermal conductivity $\xi_\kappa$ has a strong dependence on the temperature structure of the ICM. Finally, the heating (or cooling) rate due to thermal conduction is given by
\begin{equation}
	\epsilon_{\rm cond} = \frac{1}{r^2} \frac{ {\rm d} }{ {\rm d} r }\left[ r^2 F_{\rm cond} \right].
\end{equation}


\section{Evolution of the ICM}
We assume quasi-hydrostatic evolution of the ICM  \citep{Roychowdhury2004,Nath2011,Chaudhuri2011,Chaudhuri2012} such that
\begin{eqnarray}
	\frac{ {\rm d} r }{ {\rm d} M_{\rm g} } &=& \frac{1}{4 \pi r^2\rho_{\rm g}(r)} =  \frac{1}{4 \pi r^2}\left (\frac{\sigma_{\rm g}(r)}{P_{\rm g}(r)}\right )^{1/\gamma} \nonumber \\
	\frac{{\rm d} P_{\rm g} }{ {\rm d } M_{\rm g} }&=& \frac{ G M_{\rm tot}(<r) } { 4 \pi r^4 }
	\label{pressure}
\end{eqnarray}
where $\sigma_{\rm g}(r)=P_{\rm g}(r)/\rho_{\rm g}(r)^{\gamma}$ is called the entropy index  which is related to entropy of a gas ($K_{\rm g}$) as $K_{\rm g}(r)=\mu_{\rm g}\, \mu_{\rm e}^{2/3}\,m_{\rm p}^{5/3}\sigma_{\rm g}(r)$ and $M_{g}(r)$ is the gas mass enclosed up to the radius $r$. In the above equation $\gamma=5/3$ is the adiabatic index,  $m_{\rm p}$ is the mass of the proton, $\mu_{\rm g}=0.59$ and $\mu_{\rm e}=1.14$.

The ICM properties are calculated by solving Eq.~\ref{pressure} in time steps  of $\Delta t$ after incorporating heating, radiative cooling and conduction. The entropy index at a given radius changes  by amount
\begin{equation}
	\Delta \sigma_{\rm g}(r) = \frac{2}{3} \frac{\sigma_{\rm g}(r)}{P_{\rm g}(r)}\left[\epsilon_{\rm heat}(r) - \epsilon_{\rm cool}(r)-\epsilon_{\rm cond}(r) \right]\Delta t.
\end{equation}
In order to consider the redistribution of gas on account of heating and cooling, one much update the entropy index in  each time step with respect to the same gas mass shells as 
\begin{equation}
	\sigma_{\rm g} (M_{\rm g}) \rightarrow \sigma_{\rm g} (M_{\rm g}) + \Delta \sigma_{\rm g}(M_{\rm g}).
\end{equation}
The boundary condition for Eq.~\ref{pressure} is updated such that pressure at the gas mass shell initially at virial radius, is always equal to its initial pressure.  Since we heat the ICM up to the maximum radius of $0.3r_{\rm 500}$, we see that the boundary condition has no effect on the derived pressure profile in the inner regions where the impact of feedback is signification. The second boundary condition assumes $M_{\rm g}\approx0$ at $r\approx0$.

For numerical stability, the conduction term is integrated using time steps that satisfy the Courant condition \citep{Ruszkowski2002}
\begin{equation}
	\Delta t_{\rm cond} \le 0.5 \frac{{(\Delta r)^{2} n\, k_{\rm b}}}{\xi_{\kappa} (\gamma-1)}.
\end{equation}
Using the above time steps, we need to evolve the cluster profiles for the age of the cluster. We define the cluster formation epoch as the time when the cluster has a mass greater than $\frac{3}{4} M_{\rm vir}$ for the first time. This assumption is motivated by the results of the numerical simulations  which show that gravitational potential does not change much after the cluster assembles its $\frac{3}{4}$ of its total mass \cite{Navarro1997}. Using this definition for the epoch of cluster formation  \citet{Biman2004} found a convenient fit for the cluster age ($t_{\rm age}$), for a cluster, observed at a redshift of $z$
\begin{equation}
	t_{\rm age} =2.5 \times 10^9 {\rm yrs} \,(1+z)^{-2.6} \left(\frac{M_{\rm vir}}{10^{14}M_{\odot}}\right)^{-0.09}.
\end{equation}
We consider the AGN duty cycle, which is defined as the fraction of time the AGN heating is active (or ICM possesses radio bubbles), to be $50\%$. This value is  the lower limit of the duty cycle as found by \citet{Dunn2006, Birzan2012}. The cooling and conduction terms, on the other hand, are kept always on throughout the cluster age. 

For our analysis, we  will assume $\xi_{\kappa}=\xi_{\nu}=0.1$ as our fiducial conductivity and viscosity fractions.  In the next section, we will see the importance of conduction in distributing the heating of the ICM. In contrast, the viscosity fraction,  $\xi_{\nu}$, has a negligible impact on the heating profile since the dissipation length is highly dependent on  $\xi_{\kappa}$ through Eq.~\ref{dissipationlength}. 

The degree and extent of heating of ICM in the effervescent model are effectively controlled by two parameters $L^{\rm inj}_{\rm Eff}$ and $r_{\rm cutoff}$, while the corresponding two parameters for the acoustic model are: $L^{\rm inj}_{\rm Aco}$ and $f_{\rm-6}$. In both cases, the $L^{\rm inj}_{\rm Eff}$  or the $L^{\rm inj}_{\rm Aco}$ controls the amplitude, i.e the overall heating, and should be linked to the energy spewed out by the central super-massive back hole. Indeed, as we show later, the injected energy has a simple scaling relation with the mass of the black hole (related together by the underlying $M_{\rm BH}-M_{\rm halo}$ relation. The parameters $r_{\rm cutoff}$ and $f_{\rm-6}$ control the overall shape of the heating profiles, i.e it controls the radii beyond which heating exponentially/sharply falls. For the effervescent model, 
it is natural to assume heating cutoff parameter $r_{\rm cutoff}$ to be also $0.1r_{\rm 500}$ and $0.3r_{\rm 500}$ since both observations \citep{Iqbal2017b} and simulations \citep{Gaspari2014} shows no significant non-gravitational heating beyond $(0.1-0.3)r_{\rm 500}$. Similarly, for the acoustic model, the radial range of the feedback  can be suitably controlled by limiting the frequency spectrum. Since the perturbation by the sound wave depends on the wavelength, which is inversely proportional to the frequency, one can choose a minimum frequency such that the length scale matches $0.1r_{\rm 500}$ and $0.3r_{\rm 500}$ beyond which there should not be excess heating. In practice, the maximum wavelength needs to be less than $0.1r_{\rm 500}$   or $0.3r_{\rm 500}$ since conduction helps in propagating the heat further. We fix frequency range to be $f_{-6}= 0.05-0.20$ and $f_{-6}=0.01-0.20$  such that the extent of feedback is only up to $0.1r_{\rm 500}$ and $0.3r_{\rm 500}$ respectively. The choice of increasing the  upper cutoff in frequency leading to wavelengths less than the injection length-scale has a negligible effect on the results. Once, $r_{\rm cutoff}$ or $f_{-6}$ is fixed to achieve the radial dependence, one has to find a suitable overall amplitude $L^{\rm inj}_{\rm Eff}$ ($L^{\rm inj}_{\rm Aco}$) such that there is excess energy (entropy) up to a given radius.  Finally, we note that heating due to a single frequency of $f_{-6}=0.01$ or less  is not favored by X-ray observations.  Such frequencies correspond to density perturbations on length scales of $\geq 100$ kpc, for sound speed of $1000$ km s$^{-1}$ \citep{Fabian2016}, which would have been easily observed by Chandra or XMM-Newton if present.  However, if the acoustic heating is produced by the waves with a frequency spectrum, it would be difficult to separate out different perturbation modes through current X-ray observations.

\section{Results }
\subsection{Initial heating and cooling profiles}
We start by comparing the  initial heating and cooling profiles in the absence of any evolution of the ICM. Since the injected initial luminosity depends on the central black hole physics and is hence independent of the feedback models, we assume the same mechanical luminosity for a given cluster mass in both of the heating models.
This is shown in Fig.~\ref{fig1}, where the left-hand panel shows the cooling rate versus heating rate in the ICM  for the effervescent model  using initial \citet{Planelles2017} profile for three cluster masses: $2\times10^{14}M_{\rm \odot}$ (upper panel), $6\times10^{14}M_{\rm \odot}$ (middle panel) and $2\times10^{15}M_{\rm \odot}$ (bottom panel) at redshift $z=0$. The parameters of the heating model are roughly chosen such that heating not only balances the cooling but also produces excess energy in the cluster cores. 
One can see that the heating rate  with $r_{\rm cutoff}=0.3r_{\rm 500}$ (blue line) and $r_{\rm cutoff}=0.1r_{\rm 500}$ (red line)  produces excess energy in the cluster cores for all the three cluster masses. Increasing  $L^{\rm inj}_{\rm Eff}$ further will only increase the normalization of the heating profiles.  As expected, a higher amount of energy feedback is required for massive clusters.

\begin{figure*}
	\centering	
	\includegraphics[width=0.94\textwidth]{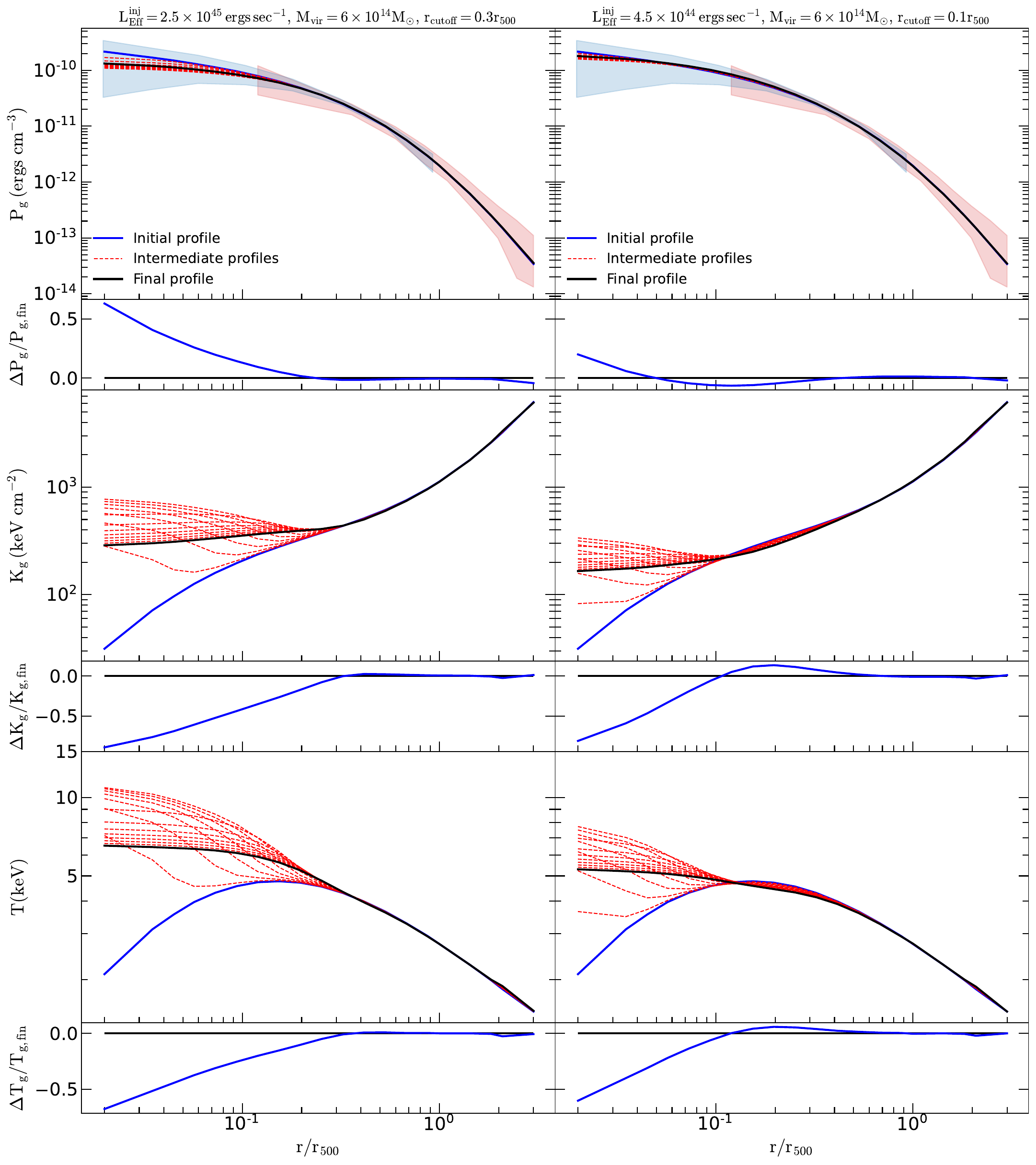}
	\caption{
		ICM thermodynamics for the effervescent model. Evolution of pressure (top), entropy (middle) and temperature (bottom) profiles as a function of radius in effervescent heating for a cluster of mass $6\times10^{14}M_{\rm \odot}$ at $z=0$ with $r_{\rm cutoff}=0.3r_{\rm 500}$ (left panel)  and  $r_{\rm cutoff}=0.1r_{\rm 500}$ (right panel) with cooling and conduction ($\xi_{\kappa}=0.1$) included. The evolution of the profiles is shown at intervals of $13\times10^8$ years with thin red dashed lines.  The total evolution time is $t_{\rm age}=2.2\times 10^{9}$ yrs and heating is turned on  during the first half of the evolution. Cooling and conduction are present throughout the evolution. The pressure (entropy) is seen to fall (rise) as the gas is heated and then rise (fall)  after the heating is switched off. Initial and final states correspond to thick solid blue and black lines respectively.  The blue and red shaded regions are the dispersion of the stacked XMM-Newton and Planck pressure profiles of 62 clusters form \citet{Planck2013V}   (their figure 4). For each panel, we also show the fractional change between the initial and final profiles ($P_{\rm g,fin}$,  $K_{\rm g,fin}$, $T_{\rm g,fin}$).
	} 
	\label{fig2}
\end{figure*}

\begin{figure*}
	\centering	
	\includegraphics[width=0.94\textwidth]{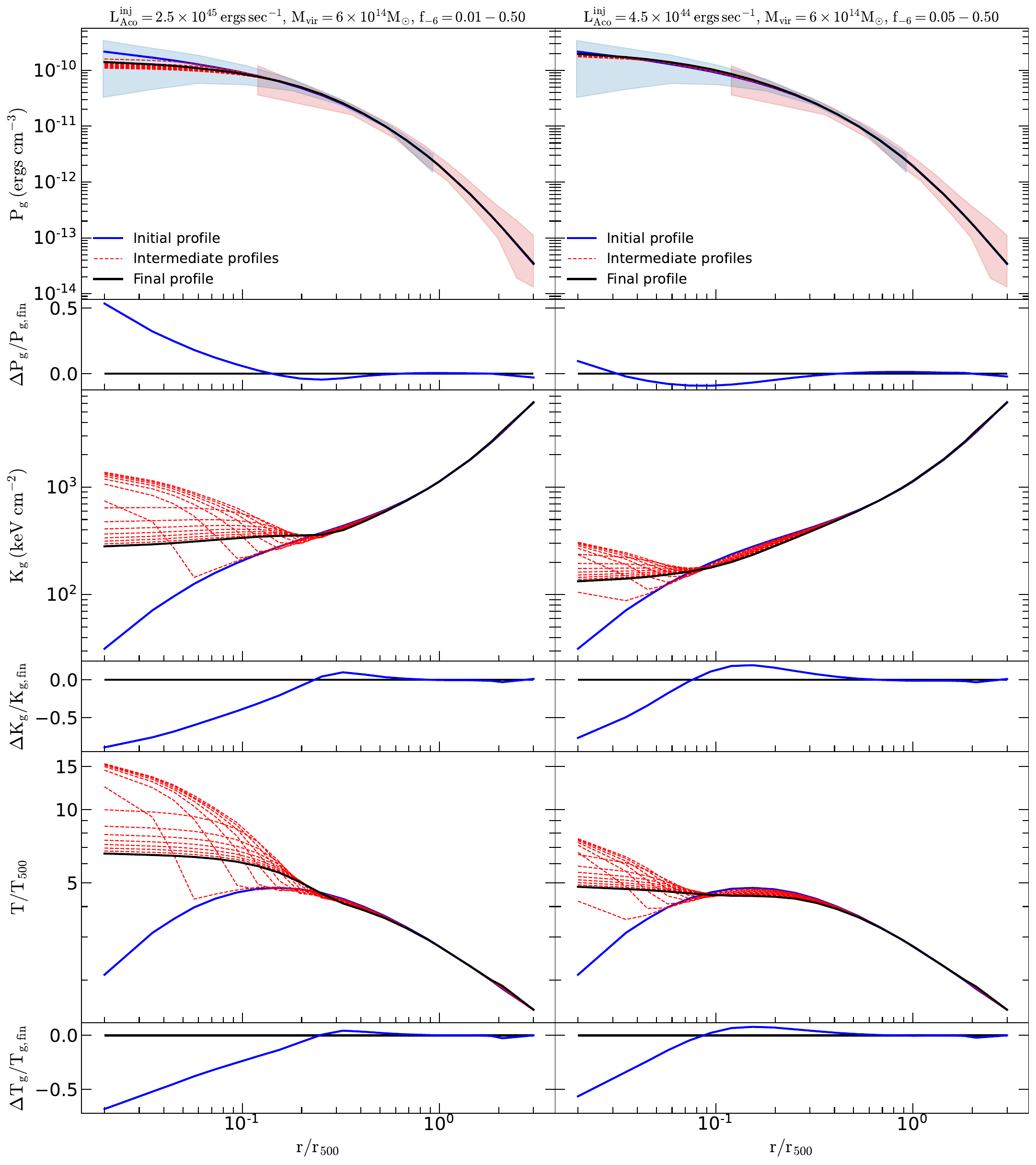}
	\caption{
		ICM thermodynamics for the acoustic model. Evolution of pressure (top), entropy (middle) and temperature (bottom) profiles as a function of radius in acoustic heating for a cluster of mass $6\times10^{14}M_{\rm \odot}$ at $z=0$ with spectrum $\zeta=1.8$ and $f_{\rm -6}=0.01-0.20$ (left panel)  and  $f_{\rm -6}=0.05-0.20$ (right panel). The evolution also includes cooling and conduction with $\xi_{\kappa}=0.1$ and  $\xi_{\nu}=0.1$. The evolution of the profiles is shown at intervals of $13\times10^8$ years with thin red dashed lines.  The total evolution time is $t_{\rm age}=2.2\times 10^{9}$ yrs and heating is turned on  during the first half of the evolution. Cooling and conduction are present throughout the evolution.
		The pressure (entropy) is seen to fall (rise) as the gas is heated and then rise (fall)  after the heating is switched off. Initial and final states correspond to thick solid blue and black lines respectively.  The blue and red shaded regions are the dispersion of the stacked XMM-Newton and Planck pressure profiles of 62 clusters form \citet{Planck2013V}   (their figure 4). For each panel, we also show the  fractional change between the initial and final profiles ($P_{\rm g,fin}$,  $K_{\rm g,fin}$, $T_{\rm g,fin}$).
	}
	\label{fig3}
\end{figure*}

\begin{figure*}
		\includegraphics[width=0.98\textwidth]{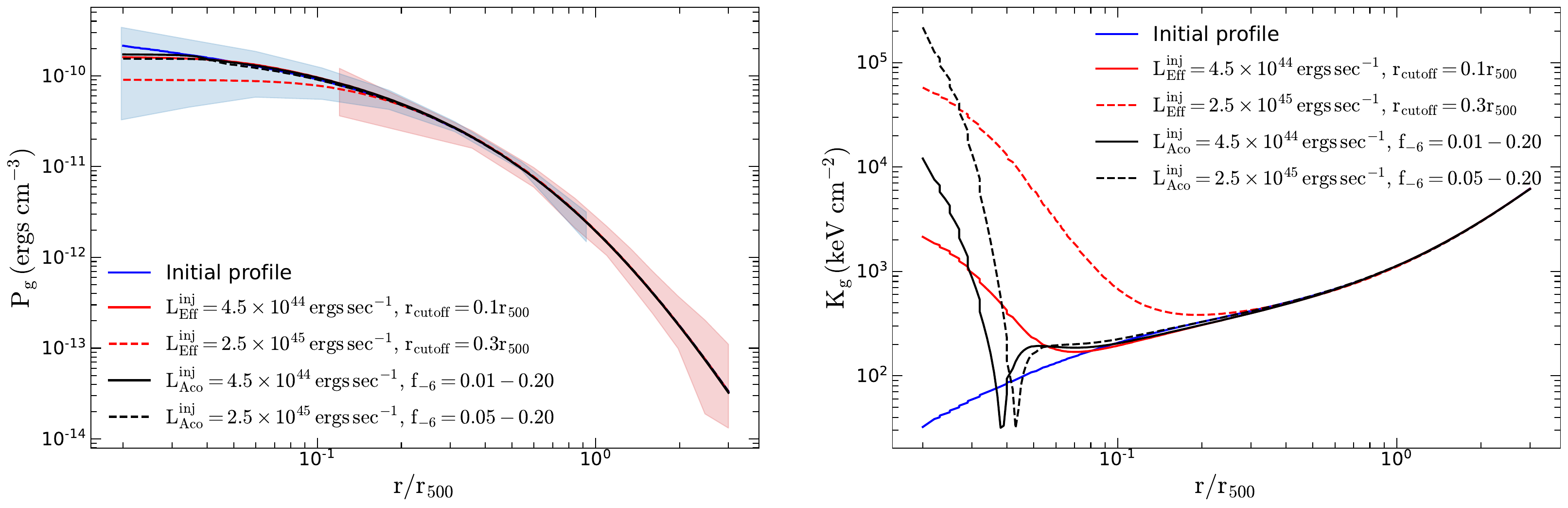}
		\caption{ Final pressure (left panel) and entropy (right panel) for effervescent and acoustic heating  without considering conduction ($\xi_{\kappa}=0$ ) for $6\times10^{14}M_{\odot}$ cluster mass. 
		}
		\label{fig4}
\end{figure*}

For the acoustic case, the right-hand panel in Fig.~\ref{fig1} shows the heating profiles for the same three cluster masses  by considering spectrum having $\zeta=1.8$ in the frequency ranges  $f_{\rm -6}=0.01-0.20$ (solid blue line) and  $f_{\rm -6}=0.05-0.20$  (solid red line), and with $\xi_{\kappa}=\xi_{\nu}=0.1$.   For comparison, we also show the heating profiles for two single frequencies, $f_{-6}=0.01$ (dashed blue line)  and $f_{-6}=0.05$ (dashed red line).  We find that for a spectrum of frequencies, the heating rate is dominated by the lowest frequency (i.e $f_{\rm -6}=0.01$ or $f_{\rm -6}=0.05$).  We note that the effective radial range of heating can be reduced  not only by increasing the  frequency but by also increasing the conductivity fraction $\xi_{\kappa}$.   As in the effervescent case, changing the magnitude of acoustic luminosity only changes the normalization of the heating profiles. 
As can be seen in the figure, the heating profiles having frequency spectrum  of  $\zeta=1.8$ with frequencies $f_{\rm-6}=0.01-0.20$ (or $f_{\rm-6}=0.05-0.20$)   produces more realistically decreasing heating profiles,  unlike in single frequency case, say, with $f_{\rm -6}=0.01$  where the heating profile is flatter in the inner region and then suddenly drops. This is easy to understand since a range of frequencies affects a range of length scales with an average contribution to all the scales till it reaches the lowest frequency (or highest wavelength) after which the heating falls off; on the other hand, a single frequency only has one dissipation length and the heating drastically falls beyond that radius. Therefore, we will only consider the acoustic model having a  frequency spectrum  in the rest of our calculations.  Next, we turn our attention to the evolution of thermodynamical profiles as a response to heating, cooling, and conduction.

\subsection{Evolution of the ICM with effervescent heating}
Fig.~\ref{fig2} shows the  evolution of pressure (top) and entropy (middle)  and temperature (bottom)  for the $6\times 10^{14} M_{\odot}$ cluster by considering effervescent model with $L^{\rm inj}_{\rm Eff}=2.5\times10^{45}$ ergs s$^{-1}$  and $r_{\rm cutoff}=0.3r_{500}$ (left panel) and  $L^{\rm inj}_{\rm Eff}=4.5\times10^{44}$ ergs s$^{-1}$ and $r_{\rm cutoff}=0.1r_{500}$ (right panel)  for the `entire' cluster radial range, i.e ($0.02-3)r_{\rm 500}$. The evolution also includes the  cooling and conduction  ($\xi_{\kappa}=0.1$).  The values of $L^{\rm inj}_{\rm Eff}$  are chosen so as to produce excess energy up to $0.3r_{500}$ (left panel ) and $0.1r_{500}$ (right panel).  The profiles are evolved for the time period  of $t_{\rm age}=2.2\times 10^{9}$ yrs with the AGN heating switched off at half the time interval. Cooling and conduction, on the other hand, are always present throughout the age of the cluster. 
The time steps used to evolve the ICM is taken to be $\approx 10^{4}-10^5$ years. However, in the figure, the profiles are plotted after each $\approx1.3\times10^{8}$ years  (thin dashed lines).    The initial profiles prior to heating are represented by thick solid blue lines in all the sub panels. Similarly, the final profiles at the end of $t_{\rm age}$ are shown by thick solid black lines. As the heating is turned on, the pressure profiles start to decrease  until the heating is stopped (at $0.5t_{\rm age}$), after which profiles start to rise back slowly. This is due to the fact that the gas is pushed out due to the central heating, which later falls back. We see that the pressure profiles during the evolution are always within the \citet{Planck2013V} observed dispersion. The entropy and temperature profiles, as expected,  show a reverse trend. They initially increase and then decrease after the heating is switched off.  The fractional difference between the initial and final profiles is also shown in Fig.~\ref{fig2}. One can see that the fractional difference can be more than 50\%  near the center and has a very strong central radial dependence.  We can also see that the fractional difference profiles becomes zero at $\sim 0.1r_{\rm500}$ and  $\sim 0.3r_{\rm500}$, as expected.

\subsection{Evolution of the  ICM with acoustic  heating}
Similarly, the  evolution of thermal profiles with acoustic heating, cooling, and conduction is shown in  Fig.~\ref{fig3}. We assume a frequency spectrum with $\zeta=1.8$ and $\xi_{\kappa}=\xi_{\nu}=0.1$.  We choose the same values of mechanical luminosity as that in effervescent heating.
We see that an acoustic luminosity of $L^{\rm inj}_{\rm Aco}=2.5\times10^{45}$ ergs s$^{-1}$ in the frequency range of $f_{-6}=0.01-0.20$  and an acoustic luminosity of $L^{\rm inj}_{\rm Aco}=4.5\times10^{44}$ ergs s$^{-1}$ in the frequency range of $f_{-6}=0.05-0.20$ could produces excess energy up to $0.3r_{500}$ and  $0.1r_{500}$, respectively.  Our results show that the optimal frequency range should be smaller than the frequency range of $f_{-6}=0.2-1$ as  predicted by \cite{Fabian2005} so as to produce the feedback up to $0.1r_{\rm500}$ or $0.3r_{\rm500}$.     Similar to the effervescent case, here also, the pressure (entropy) profile is pushed up (down) in the inner regions as the ICM is heated and then rises (falls) after the heating is shut off. Moreover, pressure profiles during the evolution also  lie within the observed \citet{Planck2013V}  dispersion. Similarly, the fractional change (also shown in Fig.~\ref{fig3}) can be more than 50\%  near the center in the acoustic heating.  One finds that in the case of acoustic heating, one gets sharp discontinuities  around $0.04r_{\rm 500}-0.05r_{\rm500}$ in the entropy and temperature profiles  as soon as the cluster is heated which then moves forward with time evolution.  However, as soon as heating is turned off, due to  conduction, one recovers smooth profiles at the end of evolution. It is also likely that  acoustic  heating due to frequencies $f_{-6}<0.01$ and  $f_{-6}>0.20$ are likely to be suppressed - the former range of frequencies would require a relatively large value of injected luminosity to balance the cooling  while the later range of frequencies would inject energy only into the very central region (producing very high entropy/temperature).

\subsection{Importance of conductivity }
Currently, there are no observational constraints about the level of conduction and convection, be it at a local or global scale. They could be significant or they may be totally suppressed. We find that for both heating models conductivity is crucial to produce  realistic thermal profiles.
In the Fig.~\ref{fig4}, we show the final pressure and entropy profiles for a $6\times10^{14}M_{\odot}$ cluster when conductivity is neglected (i.e, $\xi_{\kappa}=0$) for both the heating models for same values of mechanical energy as used before. As can be seen,  ignoring conductivity results in the negative gradient in entropy profiles near the cluster center which correspond to  the unreasonable central temperature ($30-100$ keV); however, pressure remains less affected.  The negative entropy gradient will set up a convection, which will also help in making entropy flat. This suggests that convection (turbulence) could be a critical process, especially if conduction is absent. However, modeling convection will also require assumptions regarding mixing length and the conclusions may be dependent on it. We plan to have a companion paper where we will have a detailed study importance of convection along with conduction.  We see that heating becomes more centrally peaked in the acoustic model compared to the effervescent heating and it becomes difficult to achieve  feedback beyond $0.05r_{500}$. In general, the impact of conduction lies in the fact that it tries to make the ICM  isothermal by transporting the large amount of energy injected near the center to the outer region which results in the entropy/temperature flattening.  This can be seen in Fig.~\ref{fig2} (effervescent model) and Fig.~\ref{fig3} (acoustic model) where the final temperature profiles  become more or less flat in the inner regions with $\xi_{\kappa}=0.1$.  In  the case of effervescent  heating, for a given $ L^{\rm inj}_{\rm Eff}$, higher values of the conductivity fraction will try to make the gas temperature uniform  more efficiently without changing the final profiles significantly. In particular, with conduction, we can have energy feedback reach to $0.3r_{\rm 500}$  even with a lower $r_{\rm cutoff}=0.1r_{\rm 500}$  and end up with similar final profiles as those obtained obtained with  $ L^{\rm inj}_{\rm Eff}=2.5\times10^{45}$ and $r_{\rm cutoff}=0.3r_{\rm 500}$. Its impact is more complex in the acoustic model where the heating also depends on the value of $\xi_{\kappa}$. One finds that ignoring conductivity  in the acoustic model, decreases the overall amplitude of the heating profiles; however, the effective heating profile extends to a larger radius, and one requires a relatively larger value of $L^{\rm inj}_{\rm Aco}$ to balance the cooling near the center. For a given $ L^{\rm inj}_{\rm Aco}$, as we increase the $\xi_{\kappa}$ from  our fiducial value of $0.1$, the heating rate becomes more centrally peaked but so does the conduction from the central  region to the outer region. This makes the final temperature profile more or less isothermal in the inner region. Moreover, we find that due to the very high central heating arising with frequency ranges $f_{-6}>0.01$, it is not possible to have reasonable feedback profiles up to $0.3r_{\rm 500}$ (especially for the high mass clusters) even though final profiles are isothermal in the inner region.
\subsection{Comparison with the observations}
\begin{figure}
	\includegraphics[width=0.48\textwidth]{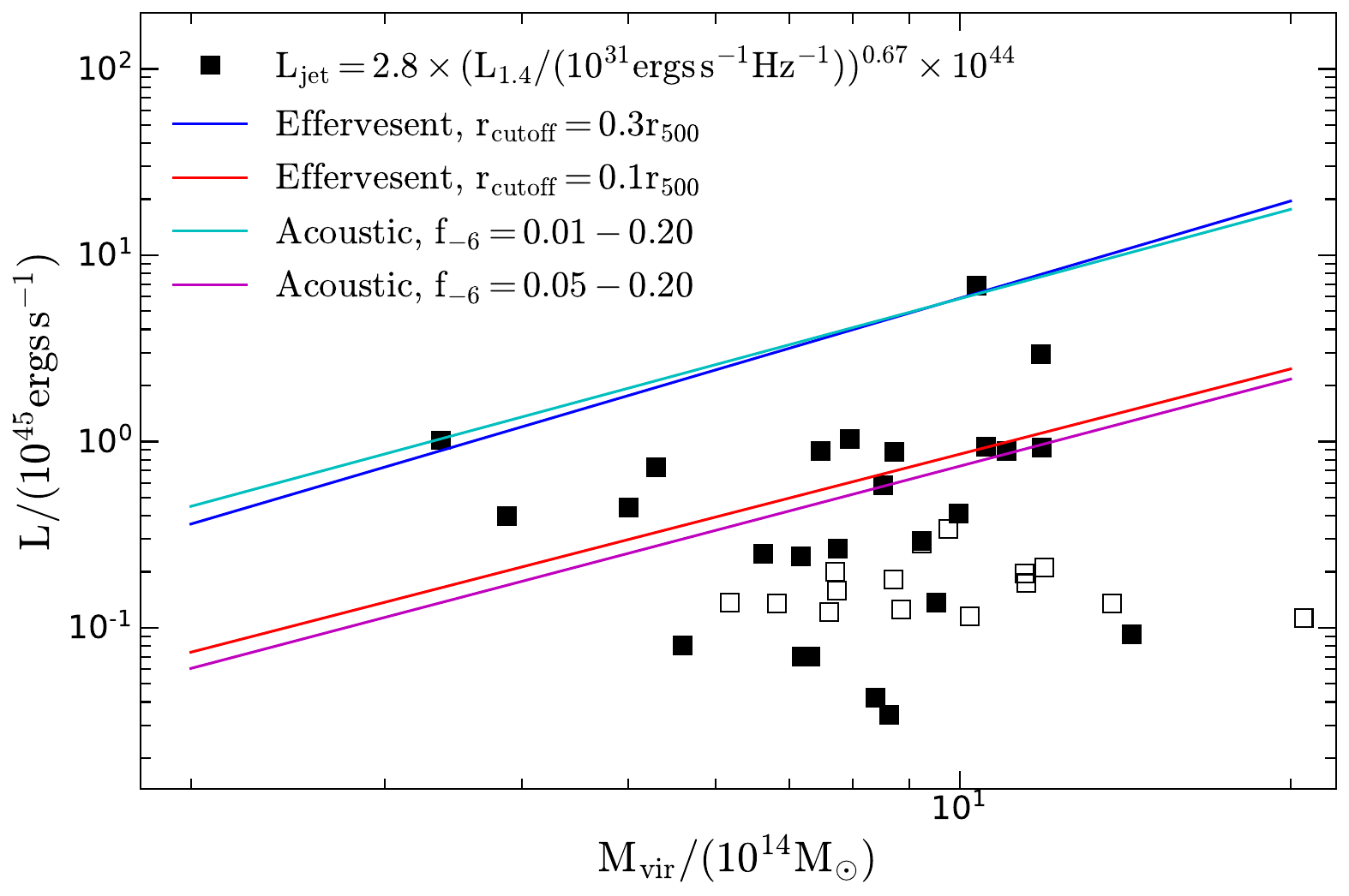}
	\caption{
		Relation between average injected luminosity and the cluster mass for different cases. Blue and red  lines are the best fit  relation for the effervescent heating for $r_{\rm cutoff} =0.3r_{\rm 500}$  and $r_{\rm cutoff} =0.1r_{\rm 500}$ respectively.  Cyan and magenta lines are the best fit  relation for the acoustic heating for frequency spectrum of $\zeta=1.8$ with $f_{\rm - 6}=0.01-0.20$  and $f_{\rm -6}=0.05-0.20$ respectively. Also shown is the expected mechanical jet power $L_{\rm jet}$ (squares) from $L_{\rm 1.4}$ measurements of \citet{Kale2015b}  using \citet{Godfrey2013}  $L_{\rm jet}-L_{\rm 1.4}$ relation for cluster of masses $\geq 10^{14} M_{\odot}$.  Open squares are based on radio upper limits. 
	}
	\label{fig5}
\end{figure}

Given that we can calculate the AGN feedback needed to balance cooling up to a certain radius for both scenarios of heating, we can estimate the mass dependence of the central injected energy. Here, we study the evolution of ICM properties of the galaxy clusters in the range $2\times10^{14}-2\times10^{15}M_{\odot}$. The Fig.~\ref{fig5} shows the derived scaling relation between the  injected effervescent (acoustic) luminosity with the total mass of the cluster such that one gets excess energy up to $0.1r_{\rm 500}$ or   $0.3r_{\rm 500}$.  For the effervescent heating and assuming $\xi_{\kappa}=0.1$  we get
\begin{eqnarray}
	\log \left(\frac{L^{\rm inj}_{\rm Eff}}{10^{45}\rm ergs\,sec^{-1}}\right)&=&-0.96+ 1.73\log \left(\frac{M_{\rm vir}}{10^{14}M_{\odot}}\right) \nonumber \\ & &\rm{for \, r_{cutoff}=0.3r_{500}}  \nonumber\\
	\log \left(\frac{L^{\rm inj}_{\rm Eff}}{10^{45}\rm ergs\,sec^{-1}}\right)&=&-1.58+ 1.52\log \left(\frac{M_{\rm vir}}{10^{14}M_{\odot}}\right) \nonumber \\ & &  \rm{for \,  r_{cutoff}=0.1r_{500}}
	\label{e:mec}
\end{eqnarray}
Similarly for the acoustic heating  and assuming $\xi_{\kappa}=\xi_{\nu}=0.1$  we get
\begin{eqnarray}
	\log \left(\frac{L^{\rm inj}_{\rm Aco}}{10^{45}\rm ergs\,sec^{-1}}\right)&=&-0.82+ 1.59\log \left(\frac{M_{\rm vir}}{10^{14}M_{\odot}}\right) \nonumber \\ & &\rm{ for \,  f_{-6}=0.01-0.20}\nonumber \\
	\log \left(\frac{L^{\rm inj}_{\rm Aco}}{10^{45}\rm ergs\,sec^{-1}}\right)&=&-1.68+ 1.55\log \left(\frac{M_{\rm vir}}{10^{14}M_{\odot}}\right) \nonumber \\ & &  \rm{for \,  f_{-6}=0.05-0.20}
	\label{e:me2}
\end{eqnarray}
Interestingly, as can be seen in the figure (and the above relations), both heating models give a similar scaling for feedback. One finds the same slope,   $L^{\rm inj}_{\rm Eff}$ (or $L^{\rm inj}_{\rm Aco}$) $\propto M_{\rm vir}^{\sim1.5}$ when heating and cooling are balanced up to $0.1r_{\rm 500}$; however, the slope for the effervescent model scaling relation is slightly steeper than acoustic model when we consider energy balance up to  a higher radius of $0.3r_{\rm 500}$.
Also plotted, in the  same figure is the estimated mechanical jet power, $L_{\rm jet}$, using the BCG radio luminosity measurements at 1.4 GHz, $L_{\rm 1.4}$,  for the cluster sample used in \citet{Iqbal2018}\footnote{$M_{\rm vir}$ is assumed to be $1.25\times M_{\rm 500}$}  (their Tab.~1)  by considering  \citet{Godfrey2013} $L_{\rm jet}-L_{\rm 1.4}$ relation for FRII galaxies and using a spectral index of 0.6
\begin{equation}
	L_{\rm jet}=2.8\times \left(\frac{L_{\rm 1.4}}{10^{31}{\rm ergs\, s}^{-1}{\rm Hz}^{-1}}\right)^{0.67}\times10^{44} \textrm{ergs } \textrm{s }^{-1}.
\end{equation}
We see that the feedback up to $0.3r_{\rm 500}$ represents the upper limit of the observed mechanical luminosity and that most of the data  centers around  $0.1r_{\rm 500}$. Note that there can be other fainter radio sources that have evaded detection but which still contribute to the heating of the ICM. 

Assuming that the scaling between the central black-hole mass and the virial mass of a galaxy halo,  $M_{\rm BH} \approx10^{9.5} (M_{\rm vir}/(10^{14}M_\odot))^{1.5}$, as given by \cite{Bandara2009} holds for cluster scales, one finds that that AGN mechanical luminosity can be approximated as  $L_{\rm inj}\approx 10^{44}$ ergs s$^{-1}$ $ M_{\rm BH}/(10^{9.5}M_\odot)$. Comparing this with the Eddington luminosity of the central SMBH, $L_{\rm Edd} \approx 10^{47.5}$ergs s$^{-1} M_{\rm BH}/(10^{9.5}M_\odot)$,
one can see that the fraction of the total luminosity available as the AGN mechanical luminosity at $0.02r_{500}$ is given by
$\epsilon_{\rm inj}=L_{\rm inj}/L_{\rm Edd} \approx 10^{-3.5}$. This falls at the lower end of the range of values used in AGN feedback simulations for the  super-Eddington accretion in order to explain the rapid assembly of $10^9$M$_\odot$ SMBHs in the first billion years of the Universe \citep{Massonneau2022}.

\section{DISCUSSION AND CONCLUSIONS}

It is interesting to compare our results with the constraints on the total injected energy from the central AGN estimated for effervescent heating by \citet{Roychowdhury2005}. They showed  that if the heating time, $t_{\rm\scriptscriptstyle heat}$, lies between $5\times10^8$ to $5\times10^9$ years, the average jet luminosity, $L^{\rm inj}$, would vary between $5\times\,10^{44}-2\times 10^{45}$ erg sec$^{-1}$ for cluster masses ranging from $4\times10^{13}M_{\rm \odot} - 10^{15}M_{\rm \odot}$. This total heating time might include short multiple episodes of the central black hole, with bubbles consisting of relativistic plasma from earlier active phases, being spread out all through the cluster atmosphere. The authors concluded that it was possible to fit the excess entropy requirements for clusters of different masses with only one pair of  $L^{\rm inj}$ and $t_{\rm\scriptscriptstyle heat}$.  They found the total energy injected into the ICM (and hence injected luminosity) to be proportional to $M_{vir}^{~1.5}$ which agrees perfectly with our estimates.  Moreover, they also estimated that this scaling is consistent with a relation between the super-massive black hole mass ($M_{\rm bh}$) in the central AGN and the cluster mass, $M_{\rm bh} \sim 10^{-5} M_{\rm vir}$, if the efficiency of conversion of energy by the accreting black hole is $\sim 0.25$. This scaling is reminiscent of the relation between black hole mass and galaxy mass \citep{Bandara2009}. 

It is worthwhile to point out that the robustness of these heating models rests upon the fact that the injected luminosity from the central AGN, be it effervescent or acoustic heating or a combination of both, lies in a similar range as demonstrated in this work. It is only natural to propose that both effervescent and acoustic heating are occurring in tandem with thermal conduction playing an important role in distributing the heat to the outer regions of the cluster atmosphere.  It is  also important  to note that other heating processes can be more important than the two models discussed here (for example, see \cite{Hillel2020}) and that the conclusions of the present paper hold only if we ignore the other heating mechanisms. Moreover, we emphasize that both heating models considered in this work have  several parameters, like   $L^{\rm inj}_{\rm Eff}$  ($L^{\rm inj}_{Aco}$) which is assumed constant, duration of the heating, and we assume the spherical model of heating. 

Currently, our approach  does not account for the observed diversity between CC and NCC clusters. In particular, we find that our heating model is not able to reproduce high densities and low entropy in the inner regions as found in CC clusters. Therefore,  it still remains to explain the wide variety of observed ICM properties, especially, the cool/non-cool core dichotomy observed in the population of galaxy clusters. \cite{Dubois11} found that the interaction between an AGN jet and the ICM gas that regulates the growth of the AGN's black hole, can naturally produce cool core clusters if the contribution of metals is neglected. However, as soon as metals are allowed to contribute to the radiative cooling, only the non cool core solution is produced.  Similarly,  it has also been argued that anisotropic thermal  conduction \citep{Barnes2019} or artificial conduction  \citep{Rasia2015}  which enhances the   mixing of gas might naturally explain the formation of CC and NCC clusters. An immediate  extension of the present work is to include cooling flows in the very central region since that would help us to produce cool cores if the mass accretion rate is high enough \citep{Biman2003}.  Additionally, one can also include convection in the evolution  which will also help to make the gas isothermal. Moreover,  the derived feedback profiles can be compared with precise  multi-wavelength observations of galaxy clusters in the  Cluster HEritage project \citep{CHEX-MATE_Collaboration2021} which will allow us to probe the extent of AGN feedback in the galaxy clusters.

The main focus of this work is to quantitatively compare effervescent and acoustic models of heating in the ICM. We study the evolution of ICM thermal profiles with these two models of AGN heating along with conduction and cooling in the clusters of  mass range of  $2\times10^{14}M_{\odot}-2\times10^{15}M_{\odot}$ at redshift, $z=0$ so as to produce excess energy up to $0.1r_{500}$ or  $0.3r_{500}$.  The heating can be controlled by tuning relevant parameters of the heating models. For effervescent heating, the relevant parameter is the outer radial cutoff of  the heating, and for acoustic heating, it is the frequency of the plasma waves. We find that for acoustic heating to work in the range $0.1r_{\rm 500}$ - $0.3r_{\rm 500}$, the optimal frequencies should lie in the range of $f_{-6}=0.01-0.20$. 
We find that one additionally requires conduction which significantly influences the properties of the ICM. As a result of the conduction, injected heat flows from the innermost regions of the cluster to the outer regions thus erasing strong temperature gradients. We also estimate the relation between the injected luminosity required to match the observations and the cluster mass.  We find that both effervescent and acoustic  produce the same scaling relations thus making it difficult to disentangle the heating models  with the X-ray and SZ observations. We find injected luminosity scales with cluster mass as $M_{\rm vir}^{\sim1.5}$ for both effervescent and acoustic heating.  Moreover, the inferred correlation is consistent  with the observed mechanical jet power and radio luminosity relation, reinforcing the idea that AGNs provide the most dominant heating in the ICM. 

To conclude, It has been shown that the power spectrum of density/pressure fluctuations in the ICM can help us to probe the AGN feedback in galaxy clusters \citep{Churazov2012,Gaspari2014b,Khatri2016,Zhuravleva2016}. Effervescent heating is expected to be associated with density fluctuations (in the form of X-ray cavities caused by bubbles) whereas acoustic heating is mainly related to pressure fluctuations. Accurate measurements of small-scale perturbations are expected from future X-ray satellites such as Athena. This will give us the ability to measure the fluctuations down to a few kpc allowing us to study the  relative contribution of AGN feedback models in heating the ICM.

\section{DATA AVAILABILITY}
The datasets generated during and/or analyzed  in this study are available upon request from the corresponding author. 

\section*{Acknowledgements}
This work was supported by CNES. AI would like to thank Raman Research Institute, Bangalore and Tata Institute of Fundamental Research, Mumbai for the support during the initial stage of this work. AI would also like to thank  Gabriel Pratt and Monique Arnaud for the useful discussions on ICM evolution.  SM acknowledges support of the Department of Atomic Energy, Government of India, under project no. 12-R\&D-TFR-5.02-0200. We would sincerely like to thank the reviewer, Noam Soker, for his insightful feedback that helped to improve the clarity of this work.


\bibliographystyle{mn2e}

\begin{thebibliography}{}
	\footnotesize{ 
		\bibitem[\protect\citeauthoryear{Adam et al.}{2015}]{Adam2015} Adam R., Comis B., Mac\'ias-P\'erez J. -F., Adane A et al., 2015, A\&A, 576, A12
		\bibitem[\protect\citeauthoryear{Andrade-Santos et al.}{2021}]{Andrade-Santos2021} Andrade-Santos, F., Pratt, G. W. et al., 2021,  A\&A, 914, 58
		\bibitem[\protect\citeauthoryear{Arnaud et al.}{2010}]{11}Arnaud M., Pratt G. W., Piffaretti R., B{\"o}hringer H., Croston J. H., Pointecouteau E., 2010, A\&A, 517, A92	
		\bibitem[\protect\citeauthoryear{Babul et~al.}{2002}]{babuletal02}Babul A.,  Balogh M.~L.,  Lewis G.~F.,  Poole G.~B.,  2002, MNRAS, 330, 329 	 
		\bibitem[\protect\citeauthoryear{Bandara et~al.}{2009}]{Bandara2009}Bandara, K.,  Crampton, D., Simard, L. , 2009, ApJ, 704, 1135
		\bibitem[\protect\citeauthoryear{Barnes et al.}{2019}]{Barnes2019}Barnes D. J. et al., 2019, MNRAS, 488, 3003
		
		\bibitem[\protect\citeauthoryear{Begelman}{2001}]{Begelman2001} Begelman, M. C. 2001, in ASP Conf. Ser. 240, Gas and Galaxy Evolution, ed. Hibbard J. E., Rupen M. P. ,  van Gorkom  J. H.  (San Francisco: ASP), 363.
		\bibitem[\protect\citeauthoryear{Battaglia et al.}{2012}]{Battaglia2012} Battaglia N., Bond J. R., Pfrommer C., Sievers J. L., 2012, ApJ, 758, 74
		\bibitem[\protect\citeauthoryear{Biffi \& Valdarnini}{2015}]{Biffi2015}Biffi V., Valdarnini R., 2015, MNRAS, 446, 2802
		\bibitem[\protect\citeauthoryear{{Binney} \& {Tabor}}{1995}]{binney&tabor95} Binney J.,  Tabor G.,  1995, MNRAS, 276, 663
		\bibitem[\protect\citeauthoryear{B\^{i}rzan et al.}{2012}]{Birzan2012}B\^{i}rzan L. et al., 2012, MNRAS, 427, 3468
		\bibitem[\protect\citeauthoryear{Bryan \& Norman}{1998}]{Bryan1998} Bryan G. L., Norman, M. L., 1998, ApJ, 495, 80
		\bibitem[\protect\citeauthoryear{Chaudhuri \& Majumdar }{2011}]{Chaudhuri2011} Chaudhuri A., Majumdar S., 2011, ApJL, 728, 41 
		\bibitem[\protect\citeauthoryear{Chaudhuri et al.}{2012}]{Chaudhuri2012} Chaudhuri A., Nath B. B., Majumdar S., 2012, ApJ, 759, 5 
		\bibitem[\protect\citeauthoryear{Chaudhuri et al.}{2013}]{Chaudhuri2013} Chaudhuri A., Majumdar S., Nath B. B. , 2013, ApJ, 776, 84
		\bibitem[\protect\citeauthoryear{CHEX-MATE Collaboration}{2021}]{CHEX-MATE_Collaboration2021} CHEX-MATE Collaboration: M. Arnaud, S. Ettori, G.W. Pratt, M. Rossetti, D. Eckert, F. Gastaldello, R. Gavazzi, S.T. Kay et al., 2021, A\&A,  650, 104
		
		
		\bibitem[\protect\citeauthoryear{Churazov et al.}{2001}]{Churazov2001} E. Churazov, M.  Br\"{u}ggen , C. R. Kaiser, H. B\"{o}hringer,  W. Forman, 2001, ApJ, 554, 261 
		\bibitem[\protect\citeauthoryear{Churazov et al.}{2012}]{Churazov2012}Churazov E., Vikhlinin A., Zhuravleva, I. et al.  2012, MNRAS,  421, 1123
		\bibitem[\protect\citeauthoryear{Duffy et al.}{2008}]{Duffy2008} Duffy A. R., Schaye J., Kay S. T., Dalla V. C.2008, MNRAS, 390, L64
		\bibitem[\protect\citeauthoryear{Dubois et~al.}{2010}]{duboisetal10} Dubois Y.,  Devriendt J.,  Slyz A., Teyssier R.,  2010, MNRAS, 409, 985
		\bibitem[\protect\citeauthoryear{Dubois et al.}{2011}]{Dubois11} Dubois Y. et al., 2011, MNRAS, 417, 1853
		\bibitem[\protect\citeauthoryear{Dunn \& Fabian }{2006}]{Dunn2006}Dunn R. J. H.,  Fabian A. C., 2006, MNRAS, 373, 959
		\bibitem[\protect\citeauthoryear{{Fabian} \& {Daines}}{{Fabian} \&   {Daines}}{1991}]{fabian&daines91} Fabian A.~C.,  Daines S.~J.,  1991, MNRAS, 252, 17P
		\bibitem[\protect\citeauthoryear{Fabian}{1994}]{Fabian_94} Fabian, A.~C., 1994, ARAA, 32, 277
		\bibitem[\protect\citeauthoryear{Fabian et al.}{2005}]{Fabian2005}Fabian A. C., Reynolds C. S., Taylor G. B., Dunn R. J. H.  2005, MNRAS, 363, 891
		
		\bibitem[\protect\citeauthoryear{Fabian et al.}{2016}]{Fabian2016}Fabian A. C.,  Walker S. A., Russell,H. R., Pinto C.,  Sanders J. S., Reynolds C. S. 2016, MNRASL, 464, 1
		
		\bibitem[\protect\citeauthoryear{{Fabjan}, {Borgani}, {Tornatore}, {Saro}, {Murante} \& {Dolag}}{{Fabjan} et~al.}{2010}]{fabjanetal10} Fabjan D.,  Borgani S.,  Tornatore L.,  Saro A.,  Murante G., Dolag K.,  2010, MNRAS, 401, 1670
		\bibitem[\protect\citeauthoryear{Gaspari et al.}{2014}]{Gaspari2014}Gaspari M., Brighenti F., Temi P., Ettori S., 2014, ApJL, 783, L10
		\bibitem[\protect\citeauthoryear{Gaspari et al.}{2014b}]{Gaspari2014b}	Gaspari M.,  Churazov E.,  Nagai D.,  Lau E. T.,  Zhuravleva I., 2014, A\&A,  569, 67
		\bibitem[\protect\citeauthoryear{Godfrey \&  Shabala}{2013}]{Godfrey2013}Godfrey  L. E. H.,  Shabala S. S., 2013, ApJ, 767, 12
		 \bibitem[\protect\citeauthoryear{Hillel \& Soker}{2020}]{Hillel2020} Hillel S., Soker, N. , 2020, ApJ, 896, 104
		\bibitem[\protect\citeauthoryear{Holder}{2001}]{Holderb2001} Holder G. P., Carlstrom J. E., 2001, ApJ, 558, 515
		
		\bibitem[\protect\citeauthoryear{Iqbal et al.}{2017a}]{Iqbal2017a}Iqbal A., Majumdar S., Nath B. B., Ettori S., Eckert D.,  Malik, M. A., 2017, 
		MNRASL, 465, L99
		\bibitem[\protect\citeauthoryear{Iqbal et al.}{2017b}]{Iqbal2017b}Iqbal A., Majumdar S., Nath B. B., Ettori S., Eckert D.,  Malik, M. A., 2017, MNRAS, 472, 713
		\bibitem[\protect\citeauthoryear{Iqbal et al.}{2018}]{Iqbal2018}Iqbal A., Nath B. B., Majumdar S., 2018, MNRASL, 480, L68
		
		\bibitem[\protect\citeauthoryear{Kale et al.}{2015}]{Kale2015b}Kale R., Venturi T., Cassano R., Giacintucci S., Bardelli S., Dallacasa D., Zucca E. 2015, A\&A, 581, 23
		\bibitem[\protect\citeauthoryear{Kaiser}{1986}]{Kaiser1986}Kaiser N., 1986, MNRAS, 222, 323
		\bibitem[\protect\citeauthoryear{{Khalatyan}, {Cattaneo}, {Schramm},{Gottl{\"o}ber}, {Steinmetz} \& {Wisotzki}}{{Khalatyan} et~al.}{2008}]{khalatyanetal08} Khalatyan A.,  Cattaneo A.,  Schramm M.,  Gottl{\"o}ber S., Steinmetz M.,    Wisotzki L.,  2008, MNRAS, 387, 13
		\bibitem[\protect\citeauthoryear{Khatri \& Gaspari.}{2015}]{Khatri2016}		Khatri R.,  Gaspari M.,  MNRAS, 2016,  463, 655
		\bibitem[\protect\citeauthoryear{McDonald et al.}{2014}]{McDonald2014}McDonald M., Benson, B. A., Vikhlinin A., Aird, K. A.  et al., 2014, ApJ, 794, 67
		\bibitem[\protect\citeauthoryear{{McCarthy}, {Schaye}, {Ponman}, {Bower}, {Booth}, {Dalla Vecchia}, {Crain}, {Springel}, {Theuns} \& {Wiersma}}{{McCarthy} et~al.}{2010}]{mccarthyetal10} McCarthy I.~G.,  Schaye J.,  Ponman T.~J.,  Bower R.~G.,  Booth C.~M.,  Dalla Vecchia C.,  Crain R.~A.,  Springel V.,  Theuns T., Wiersma R.~P.~C.,  2010, MNRAS, 406, 822
		\bibitem[\protect\citeauthoryear{Nagai et al.}{2007}]{Nagai2007}Nagai D., Kravtsov A. V.,  Vikhlinin, A., ApJ, 668, 1
		\bibitem[\protect\citeauthoryear{Nath \& Roychowdhury}{2002}]{Nath2002}Nath B. B. \&  Roychowdhury S., 2002, MNRAS, 333, 145
		\bibitem[\protect\citeauthoryear{Nath}{2003}]{Biman2003}Nath B. B., 2003, MNRASL, 339, 721
		\bibitem[\protect\citeauthoryear{Nath}{2004}]{Biman2004}Nath B. B., 2004, MNRASL, 353, 941
		\bibitem[\protect\citeauthoryear{Nath \& Majumdar}{2011}]{Nath2011}Nath B. B., Majumdar S., 2011, MNRAS, 416,271
		\bibitem[\protect\citeauthoryear{Navarro et al.}{1996}]{Navarro1996} Navarro J. F., Frenk C. S., White, S. D. M., 1996, ApJ, 462, 563
		\bibitem[\protect\citeauthoryear{Navarro et al.}{1997}]{Navarro1997} Navarro J. F., Frenk C. S., White, S. D. M., 1996, ApJ, 490, 493
		\bibitem[\protect\citeauthoryear{Okabe et al.}{2010}]{Okabe2010} Okabe N., M. Takada, K. Umetsu, Futamase T., G. P.Smith  2010, PASJ, 62, 811 
		\bibitem[\protect\citeauthoryear{Planelles et al.}{2017}]{Planelles2017}Planelles S. Fabjan, D.,  Borgani S.,  Murante G., Rasia E. et al., 2017,  MNRAS, 467, 3827 
		\bibitem[\protect\citeauthoryear{Peebles}{1980}]{Peebles1980} Peebles P. J. E., 1980, The Large Scale Structure of the Universe, Princeton UNIV. Press, Princeton
		\bibitem[\protect\citeauthoryear{Peterson et al}{2001}]{Peterson_01} Peterson, J.~ R., et al., 2001, A\&A, 365, L104
		\bibitem[\protect\citeauthoryear{Peterson et al}{2006}]{Peterson_06} Peterson J.,~R and  Fabian A.~C., 2006, Physics Reports, 427, 1 
		\bibitem[\protect\citeauthoryear{Planck Collaboration V}{2013}]{Planck2013V}Planck Collaboration V, 2013, A\&A, 550,  A2131
		
		\bibitem[\protect\citeauthoryear{Pratt et al.}{2009}]{Pratt2009} Pratt G. W.,  Croston J. H., Arnaud M., B\"{o}hringer H., 2009, A\&A, 511,  498, 361
		\bibitem[\protect\citeauthoryear{Pratt et al.}{2010}]{10}Pratt G. W. et al., 2010, A\&A, 511, 14
		
		
		
		\bibitem[\protect\citeauthoryear{{Puchwein}, {Sijacki} \&{Springel}}{{Puchwein} et~al.}{2008}]{puchweinetal08} Puchwein E.,  Sijacki D.,    Springel V.,  2008, ApJL, 687, L53
		\bibitem[\protect\citeauthoryear{{Rasia et al.}}{2015}]{Rasia2015} Rasia E. et al. 2015, ApJL, 813, L17
		\bibitem[\protect\citeauthoryear{{Rasera} \& {Chandran}}{2008}]{rasera&chandran08}
		Rasera Y.,  Chandran B.,  2008, ApJ, 685, 105
		\bibitem[\protect\citeauthoryear{{Rephaeli} \& {Silk}}{{Rephaeli} \& {Silk}}{1995}]{rephaeli&silk95}Rephaeli Y.,  Silk J.,  1995, ApJ, 442, 91
		\bibitem[\protect\citeauthoryear{Roychowdhury et al.}{2004}]{Roychowdhury2004}Roychowdhury S., Ruszkowski M., Nath B. B., Begelman, M. C.  2004, ApJ, 615, 681
		\bibitem[\protect\citeauthoryear{Roychowdhury et al.}{2005}]{Roychowdhury2005}Roychowdhury S., Ruszkowski M., Nath B. B., 2005, ApJ, 634, 90
		\bibitem[\protect\citeauthoryear{Ruszkowski \& Begelman}{2002}]{Ruszkowski2002}Ruszkowski M., Begelman M. C., 2002, ApJ, 581, 223
		\bibitem[\protect\citeauthoryear{Sereno \& Ettori}{2015}]{Sereno2015} Sereno, M. \& Ettori, S., 2015, MNRAS, 450, 3675
		\bibitem[\protect\citeauthoryear{Sokeri}{2022}]{Soker2022} Soker, N. , 2022, Universe, 8,  483
		\bibitem[\protect\citeauthoryear{Sternberg \& Soker}{2009}]{Sternberg2009} Sternberg A. \& Soker N., 20019,  MNRAS, 395, 228
		\bibitem[\protect\citeauthoryear{Shin et al.}{2016}]{Shin2016}Shin, J., Woo, J-H, Mulchaey, J.  S., ApJS, 2016, 227, 31
		\bibitem[\protect\citeauthoryear{{Sijacki} \& {Springel}}{{Sijacki} \&{Springel}}{2006}]{sijacki&springel06} Sijacki D.,  Springel V.,  2006, MNRAS, 366, 397
		\bibitem[\protect\citeauthoryear{Stott et al.}{2012}]{Stott2012} Stott  J.  P. et al., MNRAS, 2012, 422, 2213
		\bibitem[\protect\citeauthoryear{{Teyssier}, {Moore}, {Martizzi}, {Dubois} \& {Mayer}}{{Teyssier} et~al.}{2011}]{teyssieretal11} Teyssier R.,  Moore B.,  Martizzi D.,  Dubois Y.,    Mayer L.,  2011, MNRAS, 414, 195
		\bibitem[\protect\citeauthoryear{Tozzi \& Norman}{2001}]{Tozzi2001} Tozzi P., Norman C., 2001, ApJ, 546, 63
		\bibitem[\protect\citeauthoryear{Valdarnini}{2012}]{Valdarnini2012}Valdarnini R., 2012, A\&A, 546, A45
		Venturi T., Giacintucci S., Dallacasa D., Cassano R., Brunetti G., Bardelli S., Setti G., 2008, A\&A, 484, 327 
		\bibitem[\protect\citeauthoryear{{Voigt} \& {Fabian}}{2004}]{voigt&fabian04} Voigt L.~M.,  Fabian A.~C.,  2004, MNRAS, 347, 1130
		\bibitem[\protect\citeauthoryear{Voit et al.}{2005}]{Voit2005}Voit G. M., Kay S. T., Bryan G. L., 2005, ApJ, 364, 909
		\bibitem[\protect\citeauthoryear{Massonneau et al.}{2022}]{Massonneau2022}Massonneau W.,  Volonteri M., Dubois Y., Beckmann R. S.,  2022, 	arXiv:2201.08766
		\bibitem[\protect\citeauthoryear{White \& Rees}{1978}]{White-Rees_1978}White, S.~D.~M., Rees, M.~J., 1978, MNRAS, 183, 341
		\bibitem[\protect\citeauthoryear{Yang \& Reynolds}{2016}]{Yang2016}Yang H. -Y. K., Reynolds, C. S., 2016, 829, 90
		\bibitem[\protect\citeauthoryear{Zhuravleva et al.}{2016}]{Zhuravleva2016} Zhuravleva I.,  Churazov E.,  Ar\'{e}valo, P.,  Schekochihin A. A. ; Forman W. R.  et al., 2016, MNRAS, 458, 2902
		\bibitem[\protect\citeauthoryear{Zweibel et al.}{2018}]{Zweibel2018}Zweibel  E. G. , Mirnov V. V. , Ruszkowski M., Reynolds C. S. , Yang H.-Y. K. , Fabian A. C. 
	} 
\end{thebibliography}
------------------------------------------------

\end{document}